\documentclass[aps,pra,twocolumn,10pt,showpacs,superscriptaddress]{revtex4-1}

\usepackage{amsfonts}
\usepackage{graphicx}
\usepackage{amsmath}
\usepackage{amsthm}
\usepackage[colorlinks=true, citecolor=blue, urlcolor=blue ]{hyperref}
\usepackage{graphicx}

\begin{document}

\title{Wigner-Yanase skew information and entanglement generation in quantum measurement}

\author{Manik Banik}
\email{manik11ju@gmail.com}
\affiliation{Physics and Applied Mathematics Unit, Indian Statistical Institute, 203 B. T. Road, Kolkata 700108, India.}

\author{Prasenjit Deb}
\email{devprasen@gmail.com}
\affiliation{Department of Physics and Center for Astroparticle Physics and Space Science, Bose Institute, Bidhan Nagar
Kolkata - 700091, India.}

\author{Samyadeb Bhattacharya}
\email{sbh.phys@gmail.com}
\affiliation{Physics and Applied Mathematics Unit, Indian Statistical Institute, 203 B. T. Road, Kolkata 700108, India.}

\begin{abstract}
The first step of quantum measurement procedure is known as \emph{premeasurement}, during when correlation between measuring system and measurement apparatus is established. One compelling non-classical correlation is entanglement, a useful resource for various quantum information theoretic protocols. Quantifying the amount of entanglement in the premeasurement state, therefore, seeks importance from practical ground and this is the central issue of the present paper. Interestingly, for a two-label quantum system we obtain that the amount of entanglement, measured in term of \emph{negativity}, generated in premeasurement process is actually quantified by two factors: \emph{skew information} between system's initial state and the measurement direction, which quantifies the amount of information on the values of observables not commuting with the conserved quantity of the system, and \emph{mixedness parameter} of the system's initial state.         
\end{abstract}

\maketitle

\section{Introduction}
Entanglement is one of the most interesting properties of quantum system involving more than one subsystems \cite{Schrodinger,Werner,Horodecki}. This holistic property, which involves non classical correlations among subsystems, is useful resource for many quantum processes including canonical ones: quantum cryptography \cite{Crypto}, quantum teleportation \cite{Tele}, and dense coding \cite{Dense}. It is also a necessary constituent for establishing the nonlocal behavior of quantum theory \cite{nonlocality} and hence finds applications in various device independent and measurement device independent information theoretic protocols \cite{DI-MDI}. Thus entanglement generation is important for several practical purposes.

One way to generate entanglement is the \emph{premeasurement} procedure. Quantum measurement process is an interesting area of study for a long time \cite{Wheeler,Busch}. The first step of quantum measurement procedure is \emph{premeasurement} which establishes correlation between measuring system and measurement apparatus \cite{Zurek}. So, whenever some measurement is performed on a quantum system, there is a possibility that entanglement is generated between the system and the measurement device and entanglement generation in such a way attracted interest previously \cite{Zurek,Vedral}. On the other hand, in the recent past lot of efforts have been observed in studying different types of non classical correlations \cite{Modi} and hence the quantum measurement process has drawn renewed attention in creating such correlations \cite{Streltsov,Piani}. In Ref.\cite{Streltsov} the authors have shown that a von Neumann measurement on a part of a composite quantum system unavoidably creates distillable entanglement between the measurement apparatus and the system if the state has nonzero quantum discord. In a more recent article \cite{Coles}, instead of a composite system, the authors have considered a single system and they have addressed the scenario when complementary sequential measurements have been performed on the system to generate entanglement. They have obtained a lower bound of the entanglement created by the unitary interaction (premeasurement) between the system and the measurement apparatus used to measure the complementary observables sequentially. Interestingly, this lower bound resembles a well-known entropic uncertainty relations \cite{Maassen}.

In this work, like as \cite{Coles}, we have concentrated on a single system. We give our effort in finding the amount of entanglement created between the system and the apparatus by application of a single premeasurement interaction. The system may be initially prepared in any state, either pure or mixed. We try to find whether for any measurement performed on the system, prepared in an arbitrary state, there produces entanglement? And if the entanglement is produced, how much is the amount? For the simplest case i.e. a two-label quantum system (in other word \emph{qubit} system), we show that the amount of entanglement, measured in terms of negativity \cite{Sanpera,Vidal}, is quantified by two factors: (i) the mixedness of the initially prepared state of the system \cite{Jaeger}, (ii) the Wigner-Yanase skew information between the system state and the measurement direction \cite{Wigner}. We observe the following two extreme cases: (a) if the system is initially prepared in completely  mixed density matrix, then no entanglement is created between the system and the measurement apparatus, whatever the measurement is performed; (b) on the other hand, if the system is in a eigenstate of some observable (say $\sigma_z$) then maximal entanglement is created if the measurement interaction for the complementary observables (in this case any operator lying in $x-y$ plane) is switched on.

The organization of the paper is as follows. In Section-(\ref{sec2}) we briefly describe a measurement setup. Section-(\ref{sec3}) contains a brief overview on Wigner-Yanase skew information and discussion about the mixedness of a quantum state. We present our result in Section-(\ref{sec4}).

\section{The measurement setup}\label{sec2}
Consider a quantum system $\mathcal{S}$ on which some measurement will be performed. Before this measurement the system is either in some pure state $|\psi\rangle_S^{in}\in\mathcal{H}_S$ or more generally in some mixed state $\rho_S^{in}\in\mathcal{D}(\mathcal{H}_S)$, where $|\psi\rangle_S^{in}$ is a ray vector of the Hilbert space $\mathcal{H}_S$ associated with the system and $\mathcal{D}(\mathcal{H}_S)$ is the collection of density matrix (positive and trace one) acting on $\mathcal{H}_S$. Suppose a projective measurement (PVM) $X=\{X_k\}$, with $k$ outcomes, has to be performed, where each of the $X_k$ are projectors (not necessarily rank one) sum up to identity operator acting on $\mathcal{H}_S$, i.e.
$\sum_kX_k=\mathbf{1}_{\mathcal{H}_S}$. To measure any such observable $X$ on $\mathcal{S}$ one need another system called measurement device or apparatus system denoted as $M_X$. The pointer of the apparatus must indicates different positions (or different digits) depending on the different values $k$ of the measurement $X$. More precisely, the pointer states corresponding to different outcomes must be orthogonal so that they can be perfectly distinguished. 
\begin{figure}[h!]
\centering
\includegraphics[height=3cm,width=8cm]{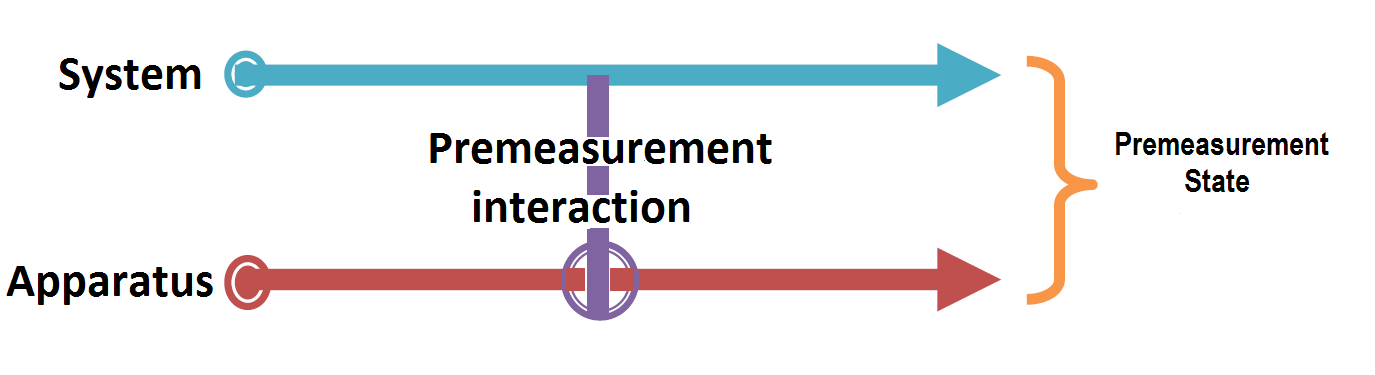}
\caption{(Color on-line) Schematic diagram describing the premeasurement procedure. Initially system and measurement apparatus are uncorrelated. Later an interaction is switched on to correlate them.}\label{fig1}
\end{figure}

The first step of a measurement process is called `premeasurement' where a correlation is established between the system and the measurement apparatus (see Fig.\ref{fig1}). Consider an orthonormal basis $\{|k\rangle_{M_X}\}$ for the apparatus system $M_X$. The interaction is made in such a way that the operator system jumps to the state $|k\rangle_{M_X}$ whenever the projector $X_k$ clicks. The required interaction can be written as
\begin{equation}\label{premea1}
|0\rangle_{M_X}|\psi\rangle_S^{in}\longrightarrow\sum_k|k\rangle_{M_X}(X_k|\psi\rangle^{in})_S=V_X|\psi\rangle_S^{in}.
\end{equation}                       
The above interaction is essentially a controlled-shift operation, and the notation is simplified by defining the isometry
\begin{equation*}
V_X:=|k\rangle_{M_X}\otimes(X_k)_S.
\end{equation*} 
In Eq.(\ref{premea1}) $|0\rangle_{M_X}$ denote the initial state of the measurement apparatus and we have consider the apparatus in a pure  state. In Ref.\cite{Vedral} the author consider the scenario where the initial state of the apparatus may be a mixed state. However, initially the system may be in a mixed state $\rho_S^{in}$. In that case the final state (also called the `premeasurement state' \cite{Zurek}) of the composite system (system $+$ apparatus) after the premeasurement is given by
\begin{equation}\label{premea2}
\rho_{SM_X}=V_X\rho_S^{in}V_X^{\dagger}
\end{equation} 

\subsection*{Hamiltonian formulation of measurement}
We have already discussed that the dynamics of any quantum measurement procedure is describe by an interaction process between the system and the measurement procedure. If some PVM $X=\{X_k\}$ is to be performed on the system then in the dynamical process of the measurement (i.e. the premeasurement process) the value of $X$ on the system state at some initial time transfer to an observable of the apparatus after the interaction. Thus the system and the apparatus become correlated. Denoting $H_S$ and $H_M$ as the system and measurement apparatus Hamiltonian respectively, the total hamiltonian can be expressed as:
\begin{equation}
H_{tot}=H_S\otimes\mathbf{1}_M+\mathbf{1}_S\otimes H_M+H_{int}, 
\end{equation}     
where $H_{int}$ represents the interaction Hamiltonian. Denote at some initial time ($t=0$) the composite state of the system and apparatus as $\rho_SM(0)=\rho_S^{in}\otimes|0\rangle_M\langle0|$. After the interaction, at time $t=\tau$ the evolved state is
\begin{equation}
\rho_{SM}(\tau)=U(\tau)\rho_{SM}(0)U^*(\tau),
\end{equation}   
where, $U(\tau):=\exp^{(-i\frac{H_{tot}\tau}{\hbar})}$. From the above expression it is clear that the composite system will preserve the product form in absence of $H_{int}$. 

Starting with arbitrary initial state of the system, in this work, our aim is to calculate the amount of entanglement between the system and the measurement apparatus in the premeasurement state for arbitrary measurement performed on the system. For two-label system we obtain an analytic expression. But before presenting our we discuss some prerequisites in the following section.

\section{Skew information and mixedness }\label{sec3}
\emph{\bf Skew information}: Quantum mechanics fundamentally differs, in many ways, from classical physics. One of the intrinsic feature of QM is Heisenberg's uncertainty principle, which says that outcomes of two noncommuting observables cannot be jointly predicted with arbitrary precision \cite{Heisenberg}. Remarkably, even a single quantum observable may display an intrinsic uncertainty as a result of the probabilistic character of quantum mechanics. There are several ways to quantify the uncertainty on single measurement. One such quantity is \emph{skew information}, introduced by Wigner and Yanase in long back 1963 \cite{Wigner}. It finds application in the study of uncertainty \cite{Luo1}, information geometry \cite{Gibilisco} and non classical correlations \cite{Luo2,Girolami}.   

If an observable $X$ is measured on the system state $\rho$, the skew information is given by
\begin{equation}
\mathcal{I}(\rho,X)=-\frac{1}{2}\mbox{Tr}[\sqrt{\rho},X]^2.
\end{equation}
TheWigner-Yanase skew information can be rewritten as
\begin{equation*}
\mathcal{I}(\rho,X)=\mbox{Tr}(\rho X^2)-\mbox{Tr}(\sqrt{\rho}X\sqrt{\rho}X)
\end{equation*}
Following are the main properties \cite{Wigner,Luo2,Lieb} of skew information:
\begin{itemize}
\item[(I)] The skew information is nonnegative and reduces to the variance if the state is pure.
\item[(II)] Skew information is convex in $\rho$ in the sense that
\begin{equation*}
\mathcal{I}(\sum\lambda_j\rho_j,X)\le\sum_j\lambda_j\mathcal{I}(\rho_j,X),
\end{equation*}
where, $\rho_j$'s are quantum state and $\lambda_j\ge0~~ \& ~~\sum_j\lambda_j=1$.
\item[(III)] For any bipartite density matrix $\rho_{ab}$ acting on tensor product Hilbert space $\mathcal{H}_a\otimes\mathcal{H}_b$ it holds that
\begin{equation*}
\mathcal{I}(\rho_{ab},X_a\otimes\mathbf{1}_b)\ge\mathcal{I}(\rho_{a},X_a),
\end{equation*} 
where, $\rho_a=\mbox{Tr}_b(\rho_{ab})$ and $\mathbf{1}_b$ is the identity operator acting on $\mathcal{H}_b$.
\end{itemize}
It has been also pointed out that skew information is the quantum version of well known Fisher information and plays important role in quantum estimation \cite{Girolami}.

\emph{\bf Mixedness of density matrix}: Not all quantum states are pure. There are also mixed quantum states represented by density matrices $\rho$ acting on the Hilbert space associated with the system. For the mixed state one has $\mbox{Tr}\rho=1$ but $\mbox{Tr}\rho^2<1$. On the other hand pure states satisfy $\mbox{Tr}\rho=\mbox{Tr}\rho^2=1$ and they are isomorphic to the ray vectors of the associated Hilbert spaces \cite{Nielsen}. The amount of mixedness of a density matrix $\rho$ is quantified as \cite{Jaeger}:
\begin{eqnarray}\label{mixed1}
\mathcal{M}(\rho)=\mbox{Tr}\rho-\mbox{Tr}\rho^2.
\end{eqnarray}  
From the above expression it is clear that mixedness is zero for pure states and maximum for the state $\frac{1}{d}\mathbf{1}_d$, where $\mathbf{1}_d$ is the identity operator acting on d-dimensional Hilbert space $\mathcal{H}_d$. The state $\frac{1}{d}\mathbf{1}_d$ is called completely mixed state. 

\section{Entanglement generation: qubit scenario}\label{sec4}
Consider any two-label quantum system generally called qubit. The Hilbert space associated with such system is the two dimensional complex vector space $\mathbb{C}^2$. The states are trace one, positive operators acting on $\mathbb{C}^2$ and can be represented as $\frac{1}{2}(\mathbf{1}_2+\vec{n}.\vec{\sigma})$, where $\vec{n}\equiv(n_x,n_y,n_z)$ is a vector in $\mathbb{R}^3$ with $|\vec{n}|^2\le 1$ and $\vec{\sigma}:=(\sigma_x,\sigma_y,\sigma_z)$ with $\sigma_i|i=x,y,z$ being Pauli matrices. The collection of all possible states for a quantum system form a convex set. For the two-label system this convex set is isomorphic to Bloch sphere with pure states lying on the surface of the sphere and mixed states inside.

Measurements on a qubit is represented by $\hat{n}.\vec{\sigma}$ with eigenstates $\frac{1}{2}(\mathbf{1}_2+\hat{n}.\vec{\sigma})$ and $\frac{1}{2}(\mathbf{1}_2-\hat{n}.\vec{\sigma})$.
\begin{figure}[h!]
\centering
\includegraphics[height=5cm,width=6cm]{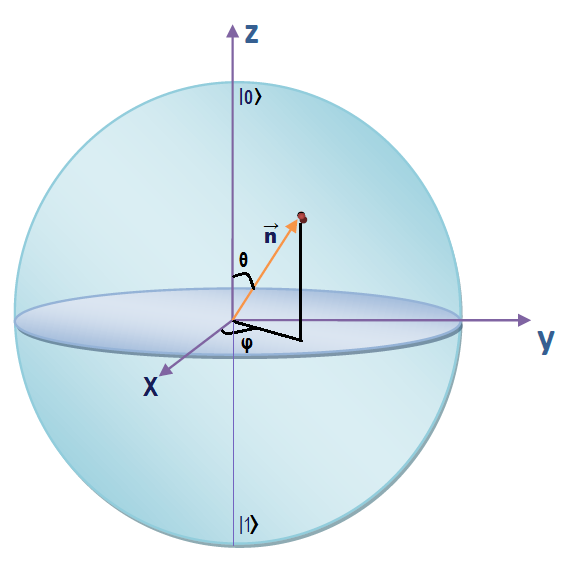}
\caption{(Color on-line) Bloch sphere representation for a two-label quantum system. The brown dot denote the initial mixed state $\rho_S^{in}$ of the system. Measurement in $\{|0\rangle,|1\rangle\}$ basis along `$Z$' direction is performed on the system.}\label{fig2}
\end{figure} 

Without any loss of generality, consider the system is initially in the state
$\rho^{in}_S=\frac{1}{2}(\mathbf{1}+\vec{n}.\vec{\sigma})$ and also consider that $\sigma_z$ measurement i.e. measurement in the basis $\{|0\rangle,|1\rangle\}$ is performed on the system (see Fig.\ref{fig2}). The unitary interaction corresponding to this measurement is given by following C-NOT operation \cite{Nielsen}:
\begin{eqnarray}\label{int}
U(|0\rangle_S\otimes|0\rangle_M)=|0\rangle_S\otimes|0\rangle_M\nonumber\\
U(|1\rangle_S\otimes|0\rangle_M)=|1\rangle_S\otimes|1\rangle_M.
\end{eqnarray}
Here $|0\rangle_M$ denotes the initial state of the apparatus.

\begin{figure}[t!]
\centering
\includegraphics[height=5cm,width=8cm]{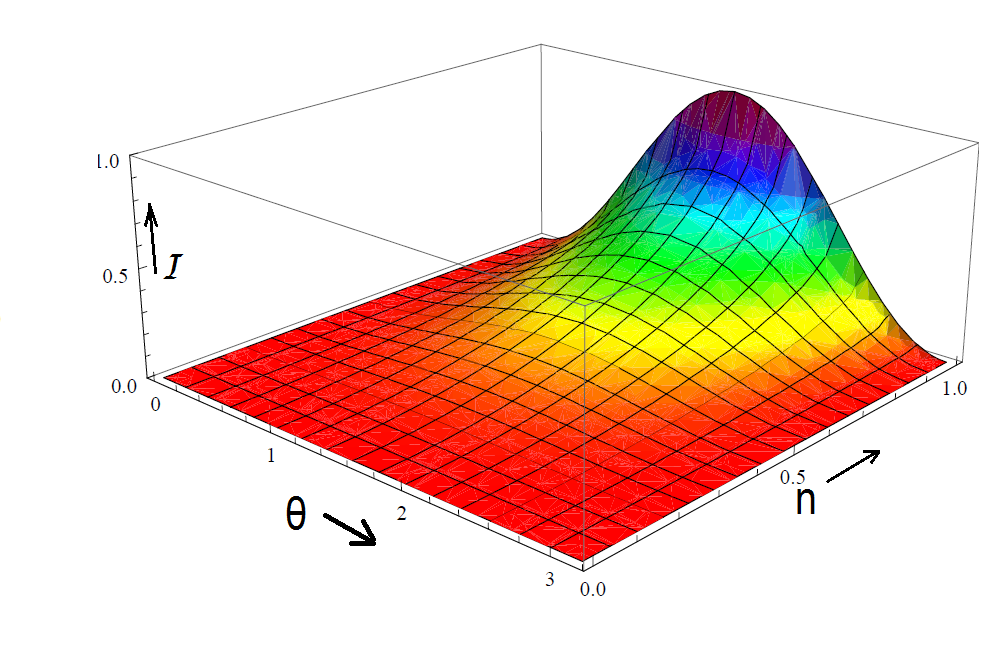}
\caption{(Color on-line) Variation of the Skew information in terms of the angle between state and measurement and the length of the state vector, as shown in Eq.(\ref{skew}).}\label{fig3}
\end{figure}

We have already discussed that the intrinsic quantum uncertainty for the measurement $\sigma_z$ on the state $\rho^{in}_s=\frac{1}{2}(\mathbf{1}_2+\vec{n}.\vec{\sigma})$ is quantified by the skew information $\mathcal{I}(\rho^{in}_S,\sigma_z)$, which we have calculated in the following. The spectral decomposition for the state $\rho^{in}_s$ reads:
\begin{eqnarray}
\rho^{in}_S&=& \frac{1}{2}(\mathbf{1}+|\vec{n}|\hat{n}.\vec{\sigma})\nonumber\\
&=&\frac{1+|\vec{n}|}{2}\frac{1}{2}(\mathbf{1}+\hat{n}.\vec{\sigma})+\frac{1-|\vec{n}|}{2}\frac{1}{2}(\mathbf{1}-\hat{n}.\vec{\sigma}).
\end{eqnarray}
Using the above spectral decomposition we can write  
\begin{equation*} \sqrt{\rho^{in}_S}=\sqrt{\frac{1+|\vec{n}|}{2}}\frac{1}{2}(\mathbf{1}+\hat{n}.\vec{\sigma})+\sqrt{\frac{1-|\vec{n}|}{2}}\frac{1}{2}(\mathbf{1}-\hat{n}.\vec{\sigma}).
\end{equation*}
Thus we have: 
\begin{eqnarray*}
\left[\sqrt{\rho^{in}_S},\sigma_z\right]&=&\left(\sqrt{\frac{1+|\vec{n}|}{2}}
-\sqrt{\frac{1-|\vec{n}|}{2}}\right)\frac{1}{2}[\hat{n}.\vec{\sigma},\sigma_z]\\
&=&\left(\sqrt{\frac{1+|\vec{n}|}{2}}-
\sqrt{\frac{1-|\vec{n}|}{2}}\right) i(n_y\sigma_x-n_x\sigma_y).
\end{eqnarray*}
The above equation further gives:
\begin{eqnarray*}
\left[\sqrt{\rho^{in}_S},\sigma_z\right]^2=-(1-\sqrt{1-|\vec{n}|^2})((n^2_x+n^2_y)\mathbf{1}_2\nonumber\\
-n_xn_y\{\sigma_x,\sigma_y\}),
\end{eqnarray*}
where, curly bracket denotes anti-commutator of two operators. Using the above expression we finally get 
\begin{eqnarray}\label{skew}
\mathcal{I}(\rho^{in}_S,\sigma_z)&=&-\frac{1}{2}\mbox{Tr}\left[\sqrt{\rho_S^{in}},\sigma_z\right]^2\nonumber\\
&=&(1-\sqrt{1-|\vec{n}|^2})(n^2_x+n^2_y).
\end{eqnarray}
Expressing the Bloch vector $\vec{n}$ in spherical coordinate as $\vec{n}\equiv(n\sin\theta\cos\phi,n\sin\theta\sin\phi,n\cos\theta)$, where $n=|\vec{n}|$ is the length of the vector and $\theta$ and $\phi$ denote the polar and azimuthal angle respectively (see Fig.\ref{fig2}), the Eq.(\ref{skew}) can be rewritten as:
\begin{equation}
\mathcal{I}(\rho^{in}_S,\sigma_z)=(1-\sqrt{1-n^2})\sin^2\theta.
\end{equation} 
From the above expression it can be conclude that for a two-label quantum system the skew information of a measurement on some state (either pure or mixed) depends on the angle between the measurement direction and the direction of the Bloch vector corresponding the quantum state and also depends on the mixedness of the density matrix (see Fig.\ref{fig3}).

Let us now quantify the mixedness of the qubit state $\rho_S^{in}=\frac{1}{2}(\mathbf{1}_2+\vec{n}.\vec{\sigma})$. We have 
\begin{eqnarray}\label{mix}
\mbox{Tr}(\rho_S^{in})^2&=&\frac{1}{4}\mbox{Tr}(\mathbf{1}_2+2\vec{n}.\vec{\sigma}+(\vec{n}.\vec{\sigma})(\vec{n}.\vec{\sigma}))\nonumber\\
&=&\frac{1}{2}(1+|\vec{n}|^2).
\end{eqnarray}
Feeding the expression of Eq.(\ref{mix}) in the definition of Eq.(\ref{mixed1}), the mixedness for a qubit state reads:
\begin{eqnarray}\label{mixed2}
\mathcal{M}(\rho_S^{in})&=&\mbox{Tr}(\rho_S^{in})-\mbox{Tr}(\rho_S^{in})^2\nonumber\\
&=&\frac{1}{2}(1-|\vec{n}|^2).
\end{eqnarray} 

According to Eq.(\ref{premea2}) after the measurement interaction as stated in Eq.(\ref{int}) the premeasurement state looks: 
\begin{eqnarray}\label{state}
\rho_{SM}&=& U\left(\frac{1}{2}(\mathbf{1}+\vec{n}.\vec{\sigma})_s\otimes|0\rangle_M \langle0|\right) U^\dagger\nonumber\\
&=& \frac{1+n_z}{2}|0\rangle_S\langle 0|\otimes|0\rangle_M\langle 0|\nonumber\\
&&+\frac{n_x+in_y}{2}|1\rangle_S\langle 0|\otimes|1\rangle_M\langle 0|\nonumber\\
&&+\frac{n_x-in_y}{2}|0\rangle_S\langle 1|\otimes|0\rangle_M\langle 1|\nonumber\\
&&+\frac{1-n_z}{2}|1\rangle_S\langle  1|\otimes|1\rangle_M\langle 1|. 
\end{eqnarray}
We are interested in the amount of entanglement between qubit system and the measurement apparatus in the state $\rho_{SM}$. There are many different measures to calculate the amount of entanglement in a state \cite{Vidal}. For our purpose we take one such well known measure called negativity \cite{Sanpera,Vidal} which is given by:
\begin{equation}
\mathcal{N}(\rho_{AB})=\frac{||\rho^{T_A}_{AB}||_1-1}{2}=\sum_{\lambda_i<0}\lambda_i,
\end{equation}
where $T_A$ denotes partial transpose with respect to the subsystem $A$, $\lambda_i$'s denote the eigenvalues of $\rho^{T_A}_{AB}$ and $||X||_1=\mbox{Tr}|X|=\mbox{Tr}\sqrt{X^{\dagger}X}$ be the trace-norm of an operator.
For the state in Eq.(\ref{state}) calculating negativity in a straightforward manner and using the expressions from  Eq.(\ref{skew}) and Eq.(\ref{mixed2}) we obtain:
\begin{eqnarray}
\mathcal{N}(\rho_{SM})&=&\left(1-\sqrt{2\mathcal{M}(\rho_S^{in})}\right)^{-\frac{1}{2}}
\frac{(\mathcal{I}(\rho^{in}_s,\sigma_z))^{\frac{1}{2}}}{2};~~|\vec{n}|\neq 0\nonumber\\
&=& 0;~~~\mbox{otherwise}. 
\end{eqnarray}
From this expression it is clear that the amount of entanglement in the premeasurement state is zero for any measurement whenever the system is initially prepared in completely mixed state (i.e $|\vec{n}|=0$). On the other hand if the system is initially prepared in some pure state lying in the $x-y$ plane (i.e. $\rho_{xy}^{in}=\frac{1}{2}(\mathbf{1}_2+\hat{n}_{xy}.\vec{\sigma})$) and the interaction for $\sigma_z$ measurement is switched on, then the premeasurement state turns out to a maximally entangled state and hence we achieved maximum entanglement. 

\section{Discussions}
Implementation of quantum measurement procedure, in laboratory, is a challenging task \cite{measurement}. Realizing the C-NOT operation deserve more attention as it has been shown that any two qubit operation can be decomposed into controlled-NOT gates between two qubits and rotations on a single qubit \cite{Sleator}. Till date different physical systems, like cold trapped ions \cite{Cirac}, cavity QED \cite{Sleator}, superconducting quantum bits \cite{Plantenberg}, have been considered for practical implementation of C-NOT operation. In this work we have quantified the amount of entanglement generated between the a two label quantum system (qubit) and the measurement apparatus when C-NOT interaction has been switched on.       

Instead of a qubit if we consider a bipartite state shared between two parties and partial von-Neumann measurement is performed on a single qubit, then also it can be shown that non-zero skew information ensures entanglement generation across the bipartition C vs AB, where C is the ancillary measurement probe. Girolami et.al have shown that for any bipartite state $\rho_{AB}$ which has non zero discord, the skew information for such state will also be non-zero \cite{Girolami}. On the other hand Streltsov et.al have shown that if any bipartite state has non-zero discord and partial von-Neumann measurement is done on a subsystem then entanglement will be generated across C vs AB cut with C being the ancillary measurement probe \cite{Streltsov}. From these two results it can be said that non zero skew information ensures entanglement generation between measurement ancillary and the bipartite quantum system. 

On the other hand, in quantum thermodynamics, work extraction problem has got a lot of attention in the recent days and it has been shown \cite{Llobet} that optimal work can be extracted from correlations present between $n$ subsystems of a multipartite quantum system. The authors have shown that for low $n$ and locally thermal subsystems, application of appropriate unitary generates a final entangled state from which maximal work can be extracted. So in work extraction problem our analysis can provide better insight as how to create a final state having maximum amount of entanglement when number of subsystem is low.

{\bf Acknowledgment:} MB like to acknowledge discussions with G. Kar, S. S. Bhattacharya,  A. Roy and A. Mukherjee. PD would like to thank DST, Govt. Of India for Financial Support.

\end{document}